\documentclass[3p,times,procedia]{elsarticle}
\usepackage{ecrc}
\usepackage[bookmarks=false]{hyperref}
    \hypersetup{colorlinks,
      linkcolor=blue,
      citecolor=blue,
      urlcolor=blue}

\volume{00}

\firstpage{1}

\journalname{Procedia Computer Science}

\runauth{Rodrigo Men\'endez et al.}

\jid{procs}

\usepackage{amssymb}

\usepackage[figuresright]{rotating}

\begin{document}
\begin{frontmatter}



\dochead{The 22nd International Conference on Mobile Systems and Pervasive Computing (MobiSPC) \\ August 4-6, 2025, Leuven, Belgium}%

\title{Next Generation Authentication for Data Spaces: An Authentication Flow Based On Grant Negotiation And Authorization Protocol For Verifiable Presentations (GNAP4VP)}


\author[a]{Rodrigo Men\'endez} 
\author[a]{Andres Munoz-Arcentales\corref{cor1}}
\author[a]{Joaqu\'in Salvach\'ua} 
\author[a]{Carlos Aparicio}
\author[a]{Irene Plaza} 
\author[a]{Gabriel Huecas}

\address[a]{Departamento de Ingeniería de Sistemas Telemáticos, Information Processing and Telecommunications Center, Universidad Polit{\'e}cnica de Madrid, Madrid 28040, Spain}

\begin{abstract}
Identity verification in Data Spaces is a fundamental aspect of ensuring security and privacy in digital environments. This paper presents an identity verification protocol tailored for shared data environments within Data Spaces. This protocol extends the Grant Negotiation and Authorization Protocol (GNAP) and integrates OpenID Connect for Verifiable Presentations (OIDC4VP) along with support for Linked Verifiable Presentations (LVP), providing a robust foundation for secure and privacy-preserving interactions. The proposed solution adheres to the principles of Self-Sovereign Identity (SSI) to facilitate decentralized, user-centric identity management while maintaining flexibility through protocol negotiation. Two alternative interaction flows are introduced: a "Wallet-Driven Interaction" utilizing OIDC4VP, and a "LVP Authorization" model for fully automated machine-to-machine communication. These flows address critical challenges encountered in Data Spaces, including privacy, interoperability, and regulatory compliance while simultaneously ensuring scalability and minimizing trust assumptions. The paper provides a detailed technical design, outlining the implementation considerations, and demonstrating how the proposed flows guarantee verifiable, secure, and efficient interactions between participants. This work contributes towards the establishment of a more trustworthy and sovereign digital infrastructure, in alignment with emerging European data governance initiatives.
\end{abstract}

\cortext[cor1]{Corresponding author.}
\end{frontmatter}

\email{joseandres.munoz@upm.es}



\section{Introduction}
In an increasingly interconnected world, digitalization has transformed all aspects of society, from personal relationships to business and administrative processes. While digitalization has brought about significant advancements, it has also introduced challenges in terms of security and trust issues that are particularly pronounced in the era of Big Data \cite{sun2021data}. 
The exponential growth of data and its central role in modern decision-making has made securing this information more critical than ever. In this context, identity management has gained fundamental importance.

The growing dependence on data has led the European Union to promote a data market with an open and secure technical infrastructure to ensure data sovereignty and trust for citizens \cite{carovano2023regulating}. Data spaces are conceived as dynamic ecosystems to facilitate the secure and transparent exchange of information based on principles of sovereignty, interoperability, and shared governance \cite{scerri2022common}. However, the conceptual framework and the principles are robust, but their implementation faces challenges in rights management, regulation, and technological development.

Within this framework, identity (understood as the set of attributes that define an individual or entity) has evolved beyond its physical-world application. It now has an intrinsic digital dimension that is indispensable for interacting with modern technologies and platforms. Digital identity has become an extension of a person’s real-world identity, and ensuring its proper verification has become a critical element for maintaining the integrity of virtual interactions \cite{robles2024digital}. 

Digital identity verification is a fundamental aspect of data management, encompassing both data spaces and broader data ecosystems \cite{rodriguez2025web}. Without trust in the authentication of digital identities, actions conducted through digital channels, ranging from routine interactions to transactions involving personal and sensitive information, become susceptible to vulnerabilities. Consequently, the reliability of digital identities serves as the cornerstone of trust in all data exchange systems.

In this context, the primary objective of this paper is to present an identity verification protocol for data spaces that strikes a harmonious balance between security, efficiency, and user-friendliness. This approach draws upon the principles of Self-Sovereign Identity (SSI) \cite{giannopoulou2023digital}, along with applicable standards and frameworks such as GNAP and OIDC4VP. These standards and frameworks facilitate the development of robust and scalable authentication systems that prioritize user privacy.

\enlargethispage{-10mm}
\section{State of the art}
Traditionally, authentication within Data Spaces has not been specifically addressed through a unified standard. While the foundational vision of Data Spaces promotes the use of Self-Sovereign Identity (SSI), conventional web-based mechanisms such as OAuth 2.0 still appear in some early-stage designs or proofs of concept \cite{incollection}. In this landscape, OIDC4VP has emerged as a bridge between SSI principles and existing authentication flows, enabling the secure, privacy-preserving presentation of Verifiable Credentials (VCs) using extensions of the OpenID Connect framework \cite{hoops2024universal}. However, these models are still rooted in the OAuth paradigm. In contrast, the Grant Negotiation and Authorization Protocol (GNAP) represents a next-generation alternative that moves away from the limitations of OAuth such as static client registration and rigid token flows. GNAP introduces more flexible, privacy-preserving authorization mechanisms that support dynamic client interactions, and fine-grained access control \cite{rfceditor9635Grant} all of which are essential for the federated and user-centric nature of Data Spaces.
 
Nevertheless, GNAP does not natively incorporate Verifiable Credentials or mechanisms specific to SSI, making it insufficient on its own for scenarios where credential-based identity verification is essential. To bridge this gap, we introduce GNAP4VP, an extension of GNAP that preserves its flexibility while adding native support for Verifiable Credential-based identity verification. Our approach extends GNAP to enable identity verification using SSI principles, leveraging OIDC4VP and Linked Verifiable Presentations for flows that require credential presentation and selective disclosure.
 
To better understand their roles and capabilities within Data Spaces, the following subsections provide an overview of the foundational protocols and technologies involved in decentralized identity verification. 

\subsection{Data Spaces}
Data Spaces represent collaborative environments where, in a simplified scenario, two distinct actors, the Consumer and the Provider, share information\cite{lycklama2022data}. These environments encompass enterprise platforms, academic repositories, open data portals, and various data ecosystems. Identity verification within such spaces presents several challenges\cite{ayala2025challenges}:
\begin{itemize}
    \item \textbf{Privacy}: Ensuring that users’ personal data is not unnecessarily exposed.
    \item \textbf{Interoperability}: Designing a system that can integrate with diverse providers and platforms.
    \item \textbf{Efficiency}: Reducing verification times without compromising security.
Regulatory
    \item \textbf{Compliance}: Adapting to regulations such as the General Data Protection Regulation (GDPR).
\end{itemize}

Addressing these aspects necessitates implementing an identity verification flow based on advanced technologies, such as Self-Sovereign Identity (SSI) \cite{shehu2024compliance, yildiz2022tutorial}.

\subsection{Authentication and Authorization Protocols}
\subsubsection{OAuth 2.0}
OAuth 2.0 is currently the most widely adopted authorization framework used to delegate access permissions in web and mobile applications. It enables users to grant third-party applications limited access to their resources without sharing their passwords, typically through access tokens issued by an Authorization Server. OAuth 2.0’s widespread adoption is due to its simplicity and compatibility with existing web technologies \cite{hardt2012rfc}.  However, despite its popularity, OAuth 2.0 suffers several critical shortcomings that impact its effectiveness across modern applications requiring dynamic, privacy-focused, and decentralized interactions. These include a rigid client registration process that hinders dynamic client interactions and fixed token issuance flows that lack flexibility for complex, adaptive use cases \cite{helmschmidt2023grant}. Furthermore, OAuth 2.0 was not originally designed with strong privacy protections, which limits its ability to support user-centric data minimization and privacy-preserving authentication required in modern identity verification scenarios.

\subsubsection{OIDC4VP}
OIDC4VP is an extension of the OpenID Connect protocol that enables the secure and privacy-preserving presentation of Verifiable Credentials (VCs) in digital interactions. Designed to align with Self-Sovereign Identity (SSI) principles, OIDC4VP allows individuals to selectively disclose credentials, such as digital IDs, certifications, or attributes to relying parties, without relinquishing control over their personal data \cite{hoops2024universal}. It builds upon existing OIDC authentication flows to support the transmission and verification of Verifiable Presentations (VPs), ensuring that the credentials are authentic, tamper-resistant, and cryptographically bound to the user. Furthermore, OIDC4VP maintains compatibility with widely adopted identity infrastructure, promoting interoperability across heterogeneous platforms while enhancing user agency and privacy. This makes it a practical bridge between traditional web identity standards and the decentralized identity paradigm \cite{ hauck2023openid, terbu2023openid}.

\subsubsection{GNAP}
GNAP is a next-generation authorization protocol designed to overcome the limitations inherent in OAuth 2.0 by providing a more flexible, dynamic, and privacy-focused framework. It supports dynamic negotiation of permissions and customizable flows that can adapt to a variety of use cases and contexts. Unlike OAuth, GNAP allows clients to interact with authorization servers without prior static registration, enabling more fluid and user-centric interactions \cite{helmschmidt2023grant}. Additionally, GNAP incorporates advanced privacy protections and data minimization strategies, making it particularly suitable for decentralized environments like Data Spaces. Its design aligns well with the federated, user-centric, and machine-to-machine interaction models that characterize modern data ecosystems, making it a promising alternative for future-proof identity verification and access control \cite{richer2020grant}. 

\subsection{Self-Sovereign Identity as the Underlying Paradigm}
Self-Sovereign Identity (SSI) is an emerging identity model that places individuals or organizations in full control of their digital identities. Unlike traditional identity systems, where identity data is stored and managed by central authorities, SSI enables users to store and present cryptographically verifiable credentials through decentralized identifiers (DIDs) and digital wallets. The core principles of SSI include user control, minimal disclosure, and interoperability. These principles align naturally with the federated and privacy-aware design of Data Spaces, where participants must be able to prove claims about themselves without relying on centralized identity providers. SSI is foundational for enabling trusted interactions in such ecosystems, making it critical to evaluate whether current authentication and authorization protocols effectively support it \cite{preukschat2021self}. 

\subsection{Verifiable Credentials and Presentations, Linked Data Proofs and Linked Verifiable Presentations}
Verifiable Credentials (VCs) are cryptographically signed statements issued by trusted entities (issuers) and held by users (holders), who can present them to service providers (verifiers) as proof of identity, qualifications, or attributes. These credentials are tamper-evident and support selective disclosure, allowing users to share only the minimum required information. Presentations of VCs—known as Verifiable Presentations (VPs)—can be constructed in a way that binds them to a specific verifier or session, protecting against credential replay or misuse.

To ensure the integrity and authenticity of VPs, SSI systems commonly employ Linked Data Proofs or other cryptographic proof mechanisms such as JSON Web Signatures (JWS). These proofs enable verifiers to validate the credentials’ origin, the holder’s control, and whether they were altered. The use of such mechanisms is essential in decentralized environments like Data Spaces, where mutual trust must be established without relying on a central trust anchor.

Linked Verifiable Presentations are a specific type of VP that leverages the Verifiable Credentials Data Model and the use of Linked Data to express credentials in a machine-readable and semantically rich way. By presenting credentials publicly in a DID Document, these presentations allow for better interoperability, extensibility, and automated reasoning across platforms. This approach is especially valuable in Data Spaces, where heterogeneous data providers and consumers must interpret identity information in a standardized and meaningful manner \cite{identityLinkedVerifiable}.

\subsection{Need for an Integrated Protocol: GNAP4VP}
To address the gaps identified above, we propose GNAP4VP, an extension of the GNAP protocol with support for Verifiable Credential-based identity verification. GNAP4VP is designed to support both dynamic authorization negotiation and privacy-preserving credential interactions, enabling participants in a Data Space to authenticate and authorize access based on SSI principles. It extends GNAP with flows and data formats defined in OIDC4VP and Linked Verifiable Presentations, allowing verifiers to request specific credentials and holders to present them securely through verifiable presentations.

Importantly, GNAP4VP is not limited to Data Spaces. Its architecture is general-purpose and can support a wide range of identity-driven interactions, including those involving everyday citizens, public administration, and private services where SSI-based verification can enhance privacy, user control, and interoperability.

To support diverse contexts of use, GNAP4VP defines two distinct interaction flows. The first, known as Wallet-Driven Interaction, is conceptually aligned with GNAP’s “Redirect-Based Interaction” but tailored for identity verification via digital wallets. Rather than redirecting the user to a third-party authorization server, control is transferred to a digital wallet—allowing users to review and approve credential sharing directly, without intermediaries.

The second flow, referred to as LVP Authorization, introduces a fully automated mechanism for credential exchange based on Linked Verifiable Presentations. While it may conceptually resemble GNAP’s “Software-Only Authorization”, where no user interaction is required, this flow significantly extends the model to meet the requirements of Self-Sovereign Identity (SSI). It enables decentralized, verifiable identity assertions using semantically rich, machine-readable formats, making it particularly suitable for machine-to-machine (M2M) scenarios in Data Spaces.

\section{Technical Overview}
\label{tech-impl}
This section presents the technical overview of the proposed GNAP4VP protocol as part of the EUNOMIA project, led by the New Generation Internet Group (GING) Research Group at Universidad Politécnica de Madrid (UPM). While the broader project focuses on developing tools and contributing to the standardization of mechanisms within the Data Spaces ecosystem, the specific work described in this paper centers on the implementation of GNAP4VP. Our aim is to demonstrate how dynamic, privacy-preserving authorization can be integrated with Verifiable Credential-based identity verification to support future interoperable and user-centric frameworks.

GNAP4VP extends GNAP, so it includes its flexible negotiation mechanism, allowing the Consumer to propose preferred flows via the initial interaction request. Importantly, verification of Verifiable Presentations follows the established standards for Data Spaces, including validating that Credentials are issued by a globally trusted authority \cite{siska2023building}. To facilitate this negotiation, the desired flow is specified in the “Grant Request” sent by the Consumer to the Provider. This petition allows the Consumer to declare support for multiple flows, while the Provider can select, accept, or reject flows based on its capabilities and policies. It is worth noting that the Consumer Machine never communicates directly with the Provider; rather, the Consumer acts as a proxy for facilitating interactions. 

In this paper, two distinct flows are proposed to support verifiable credential interactions within GNAP4VP.

\subsection{Wallet-Driven Interaction}
This flow closely resembles the “Redirect-Based Interaction” defined in GNAP, with one key difference: instead of redirecting the user to an external authorization server, control is transferred directly to the user's wallet application. This subtle change enables users to manage credential presentations, enhancing privacy and user control.

This flow can be automated in machine-to-machine scenarios, although automation is not its primary purpose. To achieve full automation, the wallet needs to support managing the entire interaction process with the verifier, especially the selection and presentation of the appropriate credentials. If automation is required but the wallet lacks this capability or cannot determine which credentials the consumer intends to present, it is advisable to use the second flow, LVP Authorization, which streamlines the process by removing these additional selection steps. 

Currently, the Wallet-Driven Interaction flow has been fully implemented and integrated within the EUNOMIA framework, demonstrating its viability for user-centric verification scenarios where human involvement is expected. This flow allows users to seamlessly review and approve credential presentations through their digital wallets, ensuring privacy and control throughout the process. However, further refinement is needed to enhance its robustness, usability, and performance before extensive testing can be conducted. Ongoing efforts focus on addressing these aspects to prepare the flow for comprehensive evaluation in diverse real-world environments.

\begin{figure}[t]
\centerline{\includegraphics[width=0.9\textwidth]{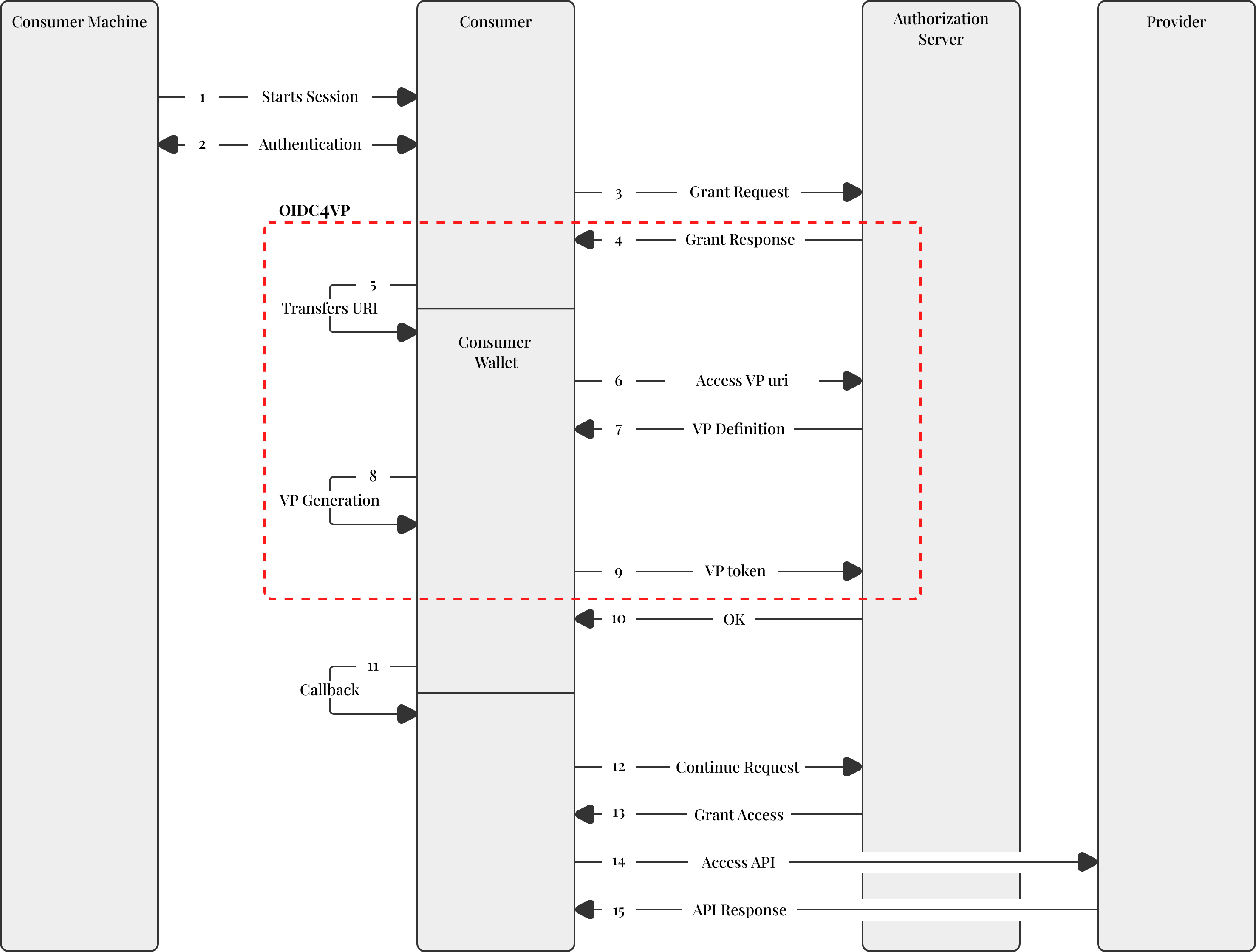}}
\label{flow-1}
\caption{Wallet-Driven Interaction Flow.}
\end{figure}

\begin{enumerate}
    \item A specific consumer machine initiates a session within its consumer global instance. \textbf{\textit{NOTE:}} \textit{There is no requirement for a consumer machine to initiate the flow; the consumer can initiate the flow itself to manage contract management or other related purposes.}
    \item The consumer machine authenticates on the consumer global instance. \textbf{\textit{NOTE:}} \textit{This aspect is beyond the scope of this paper, but each consumer can choose how to authenticate its internal machines.}
    \item The Consumer sends a Grant Request to the Provider’s authorization server, requesting access to the Provider’s API according to GNAP specifications. In this request, the Consumer also specifies its preference to use the Wallet-Driven Interaction flow. During this process, the Consumer stores the verification details associated with the redirect in the session before.
    \item The Authorization Server responds with a Grant Response, as defined in the GNAP specifications. This response provides the essential information to continue the process, including a VP Exchange URI and the continuation data linked to the request .The Authorization Server internally associates this continuation data with the ongoing request initiated by the Consumer, ensuring consistency throughout the interaction flow.
    \item The Consumer transfers the VP Exchange URI to the Consumer Wallet. This transfer can be achieved through various means, such as scanning a QR code from the wallet, copying and pasting the URI manually, or programmatically sending it to the wallet via an API request. The goal is to initiate the credential presentation process within the wallet.  \textbf{\textit{NOTE:}} \textit{The Consumer operates on behalf of each Consumer Machine, and the Verifiable Credentials used during the verification process belong to the Consumer, not the Consumer Machine.}
    \item The Wallet extracts the necessary endpoints from the VP Exchange URI and accesses them following the OIDC4VP protocol.
    \item The Authorization Server responds with the Presentation Definition required to access its protected API.
    \item Depending on how the VP Exchange URI was transferred to the Wallet, the next steps vary:
    \begin{itemize}
        \item If the URI was shared manually (e.g., via QR code scan or copy-paste), the Wallet interface will guide the user to review and select the Verifiable Credentials to be presented.
        \item If the URI was sent through an automated POST request, the Wallet must be capable of autonomously handling the entire process, including selecting appropriate credentials and generating the corresponding Verifiable Presentation (VP) without human intervention.
    \end{itemize}
    \textbf{\textit{NOTE:}} \textit{  If it is required that a machine performs the flow and cannot manage this process “LVP Authorization Flow is recommended.}
    \item The Wallet sends the VP token.
    \item The Provider validates the Verifiable Presentation (VP) token to ensure its authenticity and integrity. Once the authentication process is successfully completed, the Authorization Server finalizes the authorization of the pending request and sends the authentication result back to the Wallet, regardless of whether the flow is manual or automated. Along with this result, the Authorization Server also returns a callback URI (previously provided by the Consumer in the Grant Request), which the Wallet must use to relay the outcome of the interaction
    \begin{itemize}
        \item In manual flows, the Wallet redirects to the callback URI.
        \item In automated flows, the Wallet is expected to push the authentication results directly to the Consumer through a backend call to the provided URI.
    \end{itemize}
    \textbf{\textit{NOTE:}} \textit{ Whether the interaction should be manual or automated is indicated by the Consumer in the initial Grant Request, through the selected interaction method.}
    \item The Wallet redirects or pushes back to the Consumer URI the Provider transferred, which is the same the Consumer specified in the Grant Request. This petition also includes an interaction reference that the Authorization Server associates with the ongoing request. Additionally, some safety mechanisms are also included. \textbf{\textit{NOTE:}} \textit{ The Consumer must validate them specially in ecosystems where the Wallet is not within the boundaries of the Consumer.}
    \item The Consumer loads the continuation information and sends the interaction reference in the Continue Request.
    \item If the request is authorized, the Authorization Server grants access to the information in the form of access tokens and direct subject information to the Consumer instance.
    \item The Consumer utilizes the access tokens to invoke the Provider.
    \item The Provider validates the access token and returns the appropriate response.

\end{enumerate}

The Provider and the Consumer may subsequently engage in additional required protocols (such as contract management, data transfer, etc.).

\subsection{LVP Authorization}
The LVP Authorization flow is conceptually similar to GNAP’s Software-Only Authorization, as both are designed to enable automated, machine-to-machine (M2M) interactions without human involvement. However, the LVP Authorization flow introduces additional complexity to ensure compliance with Self-Sovereign Identity (SSI) principles and the specific verification requirements of Data Spaces. In this flow, Linked Verifiable Presentations (LVPs) are exchanged directly between the Consumer and the Authorization Server, enabling the secure and verifiable transmission of identity data. This makes the flow particularly suitable for backend services or autonomous agents that manage credentials internally and must meet high assurance levels for identity and trust.

The implementation of this flow is currently under active development, this flow aims to enable scalable, automated identity verification in decentralized environments, expanding the protocol’s applicability.

\begin{figure}[b]
\centerline{\includegraphics[width=0.9\textwidth]{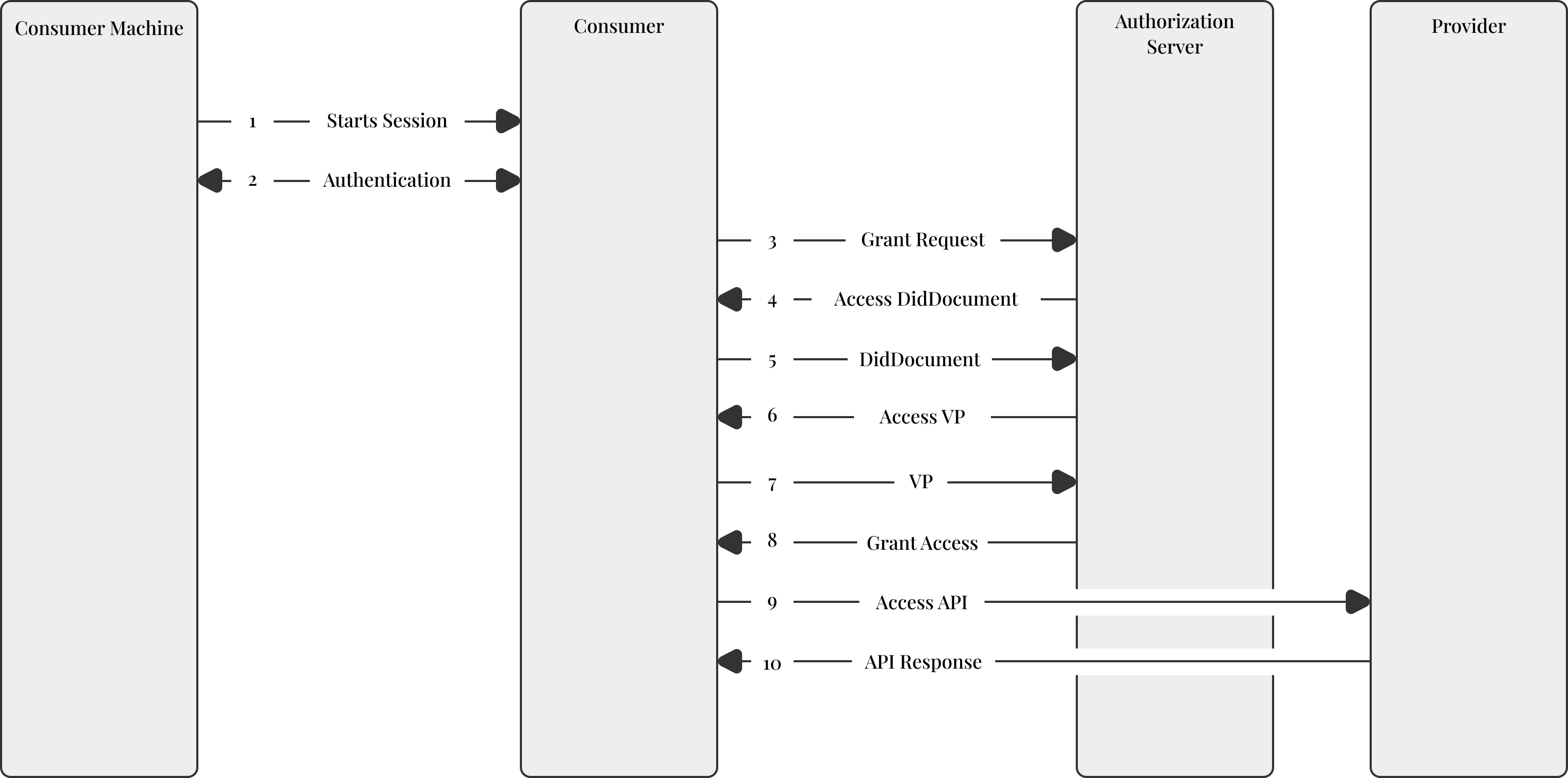}}
\label{flow-2}
\caption{LVP Authorization Flow}
\end{figure}

\begin{enumerate}
    \item A specific Consumer Machine initiates a session within its Consumer Global Instance. \textbf{\textit{NOTE:}} \textit{ As previously mentioned, there is no requirement for a Consumer Machine to initiate the flow; the Consumer can initiate the flow to manage contracts or other related purposes.}
    \item The Consumer Machine authenticates on the Consumer Global Instance. \textbf{\textit{NOTE:}} \textit{As previously mentioned, this aspect is beyond the scope of this paper, but each Consumer can determine its authentication method for internal machines.}
    \item The Consumer requests access to the Provider API following the GNAP specifications to the Authorization Server. In this instance, the Consumer does not complete the “interact” field indicating that no user interaction or redirection will occur. Instead, the “client” field is completed including signed metadata used for verification. This metadata contains information such as its validity period, the target audience, and a “did:web” identifier. \textbf{\textit{NOTE:}} \textit{ The “did:web” refers to a Decentralized Identifier (DID) that resolves to a DID Document hosted on a web server. This document contains the public keys and other verification methods required to authenticate the signature on the metadata. In this case the DID Document also presents the Linked Verifiable Presentation attribute, which references a VP to be used for identity verification.}
    \item The Provider resolves the Consumer's DID Document by querying the URL derived from the “did:web” identifier. This document contains the public key used to verify the signature of the metadata included in the request.
    \item After retrieving the DID Document, the Provider uses the verification methods specified within it to validate the metadata and the signature of the metadata. Additionally, the Provider extracts the Linked Verifiable Presentation endpoint defined in the document to retrieve the Verifiable Presentation for further validation.
    \item The Provider retrieves the Consumer’s Verifiable Presentation by accessing the endpoint previously discovered in the DID Document.
    \item After the Provider has retrieved the Verifiable Presentation, the Verifiable Presentation gets validated    to ensure its authenticity and integrity.
    \item The Authorization Server grants access to information in the form of access tokens.
    \item The Consumer utilizes the access tokens to invoke the Provider.
    \item The Provider validates the access token and returns the appropriate response.

\end{enumerate}

\section{Security Mechanisms}
\label{Security}
Security is a fundamental concern in identity verification protocols, particularly in decentralized and privacy-preserving environments such as Data Spaces. GNAP4VP integrates established security mechanisms from the GNAP framework and complements them with safeguards specific to Verifiable Credentials and Self-Sovereign Identity (SSI). Each interaction flow defined within the protocol—Wallet-Driven Interaction and LVP Authorization—relies on tailored security features that ensure the integrity, authenticity, and confidentiality of identity data throughout the exchange process.

\subsection{Wallet-Driven Interaction}
As this flow closely mirrors GNAP’s redirect-based interaction model, it inherits its core security mechanisms, such as interaction references, tokens bound to a key, and hash validation for binding the flow to an initial session. Additionally, OIDC4VP contributes features like signed presentation requests, audience restrictions, and nonces to ensure authenticity and prevent replay attacks. Together, these safeguards enable secure and privacy-preserving credential exchanges. A critical security step in this flow is also the validation of the Verifiable Credentials (VCs), ensuring that they were issued by trusted authorities.

For further details, refer to the respective GNAP and OIDC4VP specifications \cite{rfceditor9635Grant, terbu2023openid}.

\subsection{LVP Authorization}
To begin with, the initial request includes a signed metadata object inside the client field, containing information such as the validity period (validFrom, validUntil), audience, and a did:web identifier. This metadata is signed using a private key corresponding to the public key published in the DID Document resolvable via the did:web field. 
The Authorization Server retrieves and verifies the DID Document to authenticate the signature and ensure that the metadata has not been tampered. It also resolves the Linked Verifiable Presentation (LVP) endpoint included in the metadata, from which the actual Verifiable Presentation is accessed and validated. Again, a critical security step in this flow is also the validation of the Verifiable Credentials (VCs), ensuring that they were issued by trusted authorities.

To prevent replay attacks, timestamps included in the metadata define a strict validity window for the interaction. Any requests outside this window are automatically rejected. Moreover, if two requests are received with identical metadata and fall within the same validity window, the second one will be rejected as a potential replay attempt.
Additionally, the access token issued upon successful verification is cryptographically bound to the public key of the Consumer. This means that even if the token is intercepted by a third party, it cannot be used without proving possession of the corresponding private key, effectively mitigating man-in-the-middle and token theft attacks.

These mechanisms together ensure that the LVP Authorization flow maintains high assurance levels for identity and trust, even in fully automated, machine-to-machine scenarios

\section{Conclusions and Future Works}
\label{conclusions}
This paper has presented a protocol for digital identity verification tailored to the specific challenges of Data Spaces. It is grounded in Self-Sovereign Identity (SSI) principles while incorporating modern authentication standards such as GNAP and OIDC4VP. The proposed solution offers a secure, privacy-preserving, and adaptable approach to digital identity verification. The proposal accommodates both Wallet-Driven Interaction and LVP Authorization models, ensuring broad compatibility with a wide range of wallet capabilities and technical environments.

The protocol facilitates decentralized and standards-compliant verification while allowing for negotiation and customization of interaction flows. By leveraging GNAP, the protocol supports dynamic agreement between Consumers and Providers regarding the most suitable flow for each use case, enhancing interoperability and user autonomy. Furthermore, the protocol underscores the use of verifiable credentials issued by trusted authorities and emphasizes robust mechanisms for signature verification and DID resolution, thereby reinforcing security and trust.

Future work will focus on several key directions to advance and mature the proposed solution. A priority is the completion of a full reference implementation, enabling empirical validation of the protocol across diverse Data Space scenarios. While the Wallet-Driven interaction model is already implemented, further refinement is required to enhance its robustness and usability. In contrast, the LVP-based interaction model is still under development and will require full implementation and integration before it can be thoroughly evaluated. Both models will undergo practical assessments to ensure seamless operation and alignment with real-world requirements.

In parallel, a detailed evaluation of the protocol’s performance will be conducted. This includes measuring latency, scalability, and system resource consumption under various configurations and workloads, to ensure its suitability for large-scale and real-time environments.
Another important area for future exploration is enhancing interoperability with the broad spectrum of existing and emerging digital wallets. This involves developing standardized interaction patterns and testing compatibility across multiple wallet providers, ensuring seamless support for both custodial and non-custodial wallet types.

Finally, the potential for formal standardization of the GNAP4VP approach will be investigated. Contributing to existing standardization efforts and collaborating with relevant organizations will be critical to ensure long-term adoption, interoperability, and alignment with evolving best practices in decentralized identity and secure data exchange.

\section*{Acknowledgments}
  The authors would like to acknowledge the support of the EUNOMIA: "Soluciones para la soberanía, confianza y seguridad en los espacios de datos" project (ETD202300196) funded by INCIBE (Instituto de Ciberseguridad)
  \url{https://eunomia.dit.upm.es}. and also of the Agencia Estatal de Investigacion (AEI) 10.13039/501100011033, through the FuN4Date project under Grant PID2022-136684OB-C22.






\bibliographystyle{elsarticle-harv}
\bibliography{bibliography.bib}

\begin{thebibliography}{21}
\expandafter\ifx\csname natexlab\endcsname\relax\def\natexlab#1{#1}\fi
\providecommand{\url}[1]{\texttt{#1}}
\providecommand{\href}[2]{#2}
\providecommand{\path}[1]{#1}
\providecommand{\DOIprefix}{doi:}
\providecommand{\ArXivprefix}{arXiv:}
\providecommand{\URLprefix}{URL: }
\providecommand{\Pubmedprefix}{pmid:}
\providecommand{\doi}[1]{\href{http://dx.doi.org/#1}{\path{#1}}}
\providecommand{\Pubmed}[1]{\href{pmid:#1}{\path{#1}}}
\providecommand{\bibinfo}[2]{#2}
\ifx\xfnm\relax \def\xfnm[#1]{\unskip,\space#1}\fi
\bibitem[{Ayala et~al.(2025)Ayala, Bilalli, G{\'o}mez, Maz{\'o}n, Romero et~al.}]{ayala2025challenges}
\bibinfo{author}{Ayala, C.P.}, \bibinfo{author}{Bilalli, B.}, \bibinfo{author}{G{\'o}mez, C.}, \bibinfo{author}{Maz{\'o}n, J.N.}, \bibinfo{author}{Romero, O.}, et~al., \bibinfo{year}{2025}.
\newblock \bibinfo{title}{Challenges to enforce data quality in data spaces}.
\newblock \bibinfo{journal}{CEUR Workshop Proceedings} .
\bibitem[{Carovano and Finck(2023)}]{carovano2023regulating}
\bibinfo{author}{Carovano, G.}, \bibinfo{author}{Finck, M.}, \bibinfo{year}{2023}.
\newblock \bibinfo{title}{Regulating data intermediaries: The impact of the data governance act on the eu's data economy}.
\newblock \bibinfo{journal}{Computer Law \& Security Review} \bibinfo{volume}{50}, \bibinfo{pages}{105830}.
\bibitem[{Foundation(2023)}]{identityLinkedVerifiable}
\bibinfo{author}{Foundation, D.I.}, \bibinfo{year}{2023}.
\newblock \bibinfo{title}{{L}inked {V}erifiable {P}resentation}.
\newblock \bibinfo{howpublished}{\url{https://identity.foundation/linked-vp/}}.
\newblock \bibinfo{note}{[Accessed 10-04-2025]}.
\bibitem[{Giannopoulou(2023)}]{giannopoulou2023digital}
\bibinfo{author}{Giannopoulou, A.}, \bibinfo{year}{2023}.
\newblock \bibinfo{title}{Digital identity infrastructures: A critical approach of self-sovereign identity}.
\newblock \bibinfo{journal}{Digital Society} \bibinfo{volume}{2}, \bibinfo{pages}{18}.
\bibitem[{Hardt(2012)}]{hardt2012rfc}
\bibinfo{author}{Hardt, D.}, \bibinfo{year}{2012}.
\newblock \bibinfo{title}{Rfc 6749: The oauth 2.0 authorization framework}.
\bibitem[{Hauck(2023)}]{hauck2023openid}
\bibinfo{author}{Hauck, F.}, \bibinfo{year}{2023}.
\newblock \bibinfo{title}{OpenID for Verifiable Credentials: formal security analysis using the Web Infrastructure Model}.
\newblock Master's thesis. University of Stuttgart.
\bibitem[{Helmschmidt et~al.(2023)Helmschmidt, Hosseyni, K{\"u}sters, Pruiksma, Waldmann and W{\"u}rtele}]{helmschmidt2023grant}
\bibinfo{author}{Helmschmidt, F.}, \bibinfo{author}{Hosseyni, P.}, \bibinfo{author}{K{\"u}sters, R.}, \bibinfo{author}{Pruiksma, K.}, \bibinfo{author}{Waldmann, C.}, \bibinfo{author}{W{\"u}rtele, T.}, \bibinfo{year}{2023}.
\newblock \bibinfo{title}{The grant negotiation and authorization protocol: attacking, fixing, and verifying an emerging standard}, in: \bibinfo{booktitle}{European Symposium on Research in Computer Security}, \bibinfo{organization}{Springer}. pp. \bibinfo{pages}{222--242}.
\bibitem[{Hoops and Matthes(2024)}]{hoops2024universal}
\bibinfo{author}{Hoops, F.}, \bibinfo{author}{Matthes, F.}, \bibinfo{year}{2024}.
\newblock \bibinfo{title}{A universal system for openid connect sign-ins with verifiable credentials and cross-device flow}, in: \bibinfo{booktitle}{2024 IEEE International Conference on Blockchain and Cryptocurrency (ICBC)}, \bibinfo{organization}{IEEE}. pp. \bibinfo{pages}{296--298}.
\bibitem[{Imbault()}]{rfceditor9635Grant}
\bibinfo{author}{Imbault, F.}, .
\newblock \bibinfo{title}{{R}{F}{C} 9635: {G}rant {N}egotiation and {A}uthorization {P}rotocol ({G}{N}{A}{P}) --- rfc-editor.org}.
\newblock \bibinfo{howpublished}{\url{https://www.rfc-editor.org/rfc/rfc9635.html}}.
\newblock \bibinfo{note}{[Accessed 29-05-2025]}.
\bibitem[{Kuperberg and Klemens(2022)}]{incollection}
\bibinfo{author}{Kuperberg, M.}, \bibinfo{author}{Klemens, R.}, \bibinfo{year}{2022}.
\newblock \bibinfo{title}{Integration of self-sovereign identity into conventional software using established iam protocols: A survey}, in: \bibinfo{booktitle}{Open Identity Summit 2022}. \bibinfo{publisher}{Gesellschaft für Informatik e.V.}, \bibinfo{address}{Bonn}, pp. \bibinfo{pages}{51--62}.
\newblock \DOIprefix\doi{10.18420/OID2022_04}.
\bibitem[{Lycklama(2022)}]{lycklama2022data}
\bibinfo{author}{Lycklama, D.}, \bibinfo{year}{2022}.
\newblock \bibinfo{title}{Data space functionality}, in: \bibinfo{booktitle}{Designing Data Spaces: The Ecosystem Approach to Competitive Advantage}. \bibinfo{publisher}{Springer International Publishing Cham}, pp. \bibinfo{pages}{521--534}.
\bibitem[{Preukschat and Reed(2021)}]{preukschat2021self}
\bibinfo{author}{Preukschat, A.}, \bibinfo{author}{Reed, D.}, \bibinfo{year}{2021}.
\newblock \bibinfo{title}{Self-sovereign identity}.
\newblock \bibinfo{publisher}{Manning Publications}.
\bibitem[{Richer and Parecki(2020)}]{richer2020grant}
\bibinfo{author}{Richer, F.}, \bibinfo{author}{Parecki, A.}, \bibinfo{year}{2020}.
\newblock \bibinfo{title}{Grant negotiation and authorization protocol (gnap)}.
\bibitem[{Robles-Carrillo(2024)}]{robles2024digital}
\bibinfo{author}{Robles-Carrillo, M.}, \bibinfo{year}{2024}.
\newblock \bibinfo{title}{Digital identity: an approach to its nature, concept, and functionalities}.
\newblock \bibinfo{journal}{International Journal of Law and Information Technology} \bibinfo{volume}{32}, \bibinfo{pages}{eaae019}.
\bibitem[{Rodr{\'\i}guez-Doncel(2025)}]{rodriguez2025web}
\bibinfo{author}{Rodr{\'\i}guez-Doncel, V.}, \bibinfo{year}{2025}.
\newblock \bibinfo{title}{Web technologies for decentralised identity}.
\newblock \bibinfo{journal}{Governance and Control of Data and Digital Economy in the European Single Market} , \bibinfo{pages}{111}.
\bibitem[{Scerri et~al.(2022)Scerri, Tuikka, de~Vallejo and Curry}]{scerri2022common}
\bibinfo{author}{Scerri, S.}, \bibinfo{author}{Tuikka, T.}, \bibinfo{author}{de~Vallejo, I.L.}, \bibinfo{author}{Curry, E.}, \bibinfo{year}{2022}.
\newblock \bibinfo{title}{Common european data spaces: Challenges and opportunities}.
\newblock \bibinfo{journal}{Data Spaces: Design, Deployment and Future Directions} , \bibinfo{pages}{337--357}.
\bibitem[{Shehu(2024)}]{shehu2024compliance}
\bibinfo{author}{Shehu, A.S.}, \bibinfo{year}{2024}.
\newblock \bibinfo{title}{On the compliance of self-sovereign identity with gdpr principles: A critical review}.
\newblock \bibinfo{journal}{arXiv preprint arXiv:2409.03624} .
\bibitem[{Siska et~al.(2023)Siska, Karagiannis, Drobics et~al.}]{siska2023building}
\bibinfo{author}{Siska, V.}, \bibinfo{author}{Karagiannis, V.}, \bibinfo{author}{Drobics, M.}, et~al., \bibinfo{year}{2023}.
\newblock \bibinfo{title}{Building a dataspace: Technical overview}.
\newblock \bibinfo{journal}{Gaia-X Hub Austria} .
\bibitem[{Sun et~al.(2021)Sun, Zhang and Fang}]{sun2021data}
\bibinfo{author}{Sun, L.}, \bibinfo{author}{Zhang, H.}, \bibinfo{author}{Fang, C.}, \bibinfo{year}{2021}.
\newblock \bibinfo{title}{Data security governance in the era of big data: status, challenges, and prospects}.
\newblock \bibinfo{journal}{Data Science and Management} \bibinfo{volume}{2}, \bibinfo{pages}{41--44}.
\bibitem[{Terbu et~al.(2023)Terbu, Lodderstedt, Yasuda and Looker}]{terbu2023openid}
\bibinfo{author}{Terbu, O.}, \bibinfo{author}{Lodderstedt, T.}, \bibinfo{author}{Yasuda, K.}, \bibinfo{author}{Looker, T.}, \bibinfo{year}{2023}.
\newblock \bibinfo{title}{Openid for verifiable presentations-draft 18}.
\newblock \bibinfo{journal}{Internet draft} .
\bibitem[{Yildiz et~al.(2022)Yildiz, K{\"u}pper, Thatmann, G{\"o}nd{\"o}r and Herbke}]{yildiz2022tutorial}
\bibinfo{author}{Yildiz, H.}, \bibinfo{author}{K{\"u}pper, A.}, \bibinfo{author}{Thatmann, D.}, \bibinfo{author}{G{\"o}nd{\"o}r, S.}, \bibinfo{author}{Herbke, P.}, \bibinfo{year}{2022}.
\newblock \bibinfo{title}{A tutorial on the interoperability of self-sovereign identities}.
\newblock \bibinfo{journal}{arXiv preprint arXiv:2208.04692} .

\end{thebibliography}



\clearpage

\normalMode

\end{document}